\documentclass[aps,prl,twocolumn,superscriptaddress]{revtex4}
\usepackage{graphics,epsfig}

\newcommand{\beq}{\begin{equation}}
\newcommand{\eeq}{\end{equation}}
\newcommand{\bea}{\begin{eqnarray}}
\newcommand{\eea}{\end{eqnarray}}
\newcommand{\beas}{\begin{eqnarray*}}
\newcommand{\eeas}{\end{eqnarray*}}

\newcommand{\pdag}{{\phantom{\dagger}}}
\newcommand{\nn}{\nonumber}

\begin{document}


\title{Scaling and Decoherence in the Out--of--Equilibrium
Kondo Model}

\author{Stefan Kehrein} 
\affiliation{Theoretische Physik III -- Elektronische Korrelationen und
Magnetismus, Universit{\"a}t Augsburg, 86135 Augsburg, Germany} 


\date{\today}

\begin{abstract}
We study the Kondo effect in quantum dots in an out--of--equilibrium state
due to an applied dc--voltage bias. Using the method of infinitesimal
unitary transformations (``flow equations''), we develop a perturbative
scaling picture that naturally contains both equilibrium coherent and
non--equilibrium decoherence effects. This framework allows one to study the
competition between Kondo effect and current--induced decoherence,
and it establishes a large regime dominated by single--channel Kondo physics
for asymmetrically coupled quantum dots.
\end{abstract}


\maketitle




Since the first experimental observations of the Kondo effect in the Coulomb
blockade regime of quantum dots 
\cite{GoldhaberGordon98,Cronenwett98,Schmid98}, 
a wealth of experimental and theoretical 
work has addressed the properties of this highly controllable correlated
electron system. If a quantum dot weakly coupled to two leads carries a net spin, 
resonant tunneling through the dot becomes possible
and leads to a Kondoesque increase of the conductance up to the unitarity
limit upon lowering temperature \cite{Glazman88,Ng88,Wiel00}. 
These realizations of the Kondo effect in quantum dots 
have led to new questions related to the out--of--equilibrium nature of the
Kondo system with a stationary current for an applied voltage bias. 
Despite many theoretical efforts 
(e.g., Refs.
\cite{Appelbaum67,Solyom68,Koenig96,Glazman99,Rosch,Schiller95,Parcollet02}),
a satisfactory theory for the out--of--equilibrium Kondo effect
does not yet exist.
Most theoretical methods that have been developed for equilibrium \cite{Hewson}
cannot easily be generalized to the non--equilibrium situation.

In this Letter, we focus on the case of large voltage bias $V\gg T_{\rm K}$
at zero temperature ($T=0$)
as a step towards a more complete understanding of the out--of--equilibrium
Kondo model. Kaminski {\it et al.} \cite{Glazman99}
first suggested a ``poor man's'' scaling method, and
subsequently Rosch {\it et al.} \cite{Rosch}
developed a more sophisticated approach
based on frequency--dependent vertices and Keldysh diagrammatics. Both groups
noted that decoherence generated by the current is essential
since it introduces the decoherence rate
$\Gamma_{\rm rel}\propto V/\ln^2(V/T_{\rm K})$ due to non--equilibrium spin relaxation processes.
However, a scaling picture in a 
Hamiltonian framework in which this decoherence scale emerges naturally
has as yet not been developed.
This Letter shows that the method of infinitesimal unitary transformations
(``flow equations'') 
\cite{Wegner94,Wilson93}
provides such a suitable generalization of Anderson's ``poor man's'' 
scaling picture \cite{Anderson70}. 
It allows one to 
study the flow of coupling constants and the ensuing phase diagram 
in a way that is
similar to the scaling analysis of an equilibrium
problem, and it establishes a large regime dominated by single--channel Kondo physics
for asymmetrically coupled quantum dots.

We consider a Hamiltonian describing a spin-1/2 degree of freedom~$\vec S$
coupled to two leads $a,a'=l,r$ with voltage bias $V$ 
as an effective model for a quantum dot in the Kondo regime
\beq
H=\sum_{a,p,\alpha} (\epsilon^\pdag_p-\mu^\pdag_a)
c^\dag_{ap\alpha} c^\pdag_{ap\alpha}\!
+\sum_{a',a} J^\pdag_{a'a} \sum_{p',p} \vec S\, \cdot\, \vec s_{(a'p')(ap)}
\label{Kondo_nonequ}
\eeq
Here $\vec s_{(a'p')(ap)}=\frac{1}{2}\sum_{\alpha,\beta} c^\dag_{a'p'\alpha}
\vec\sigma^\pdag_{\alpha\beta} c^\pdag_{ap\beta}$,
$p,p'$
are momentum labels and $\mu_{l,r}=\pm V/2$.
If the quantum dot can be described by an Anderson impurity model
with tunneling rates $\Gamma_{l,r}$ from the left/right lead, the coupling
constants are related by $J_{lr}^2=J^\pdag_{ll} J^\pdag_{rr}$
and $J_{ll}/J_{rr}=\Gamma_l/\Gamma_r$ 
(notice $J^\pdag_{lr}=J^\pdag_{rl}$ for hermiticity) \cite{Glazman99}.

\begin{figure}[b]
\includegraphics[clip=true,width=8.0cm]{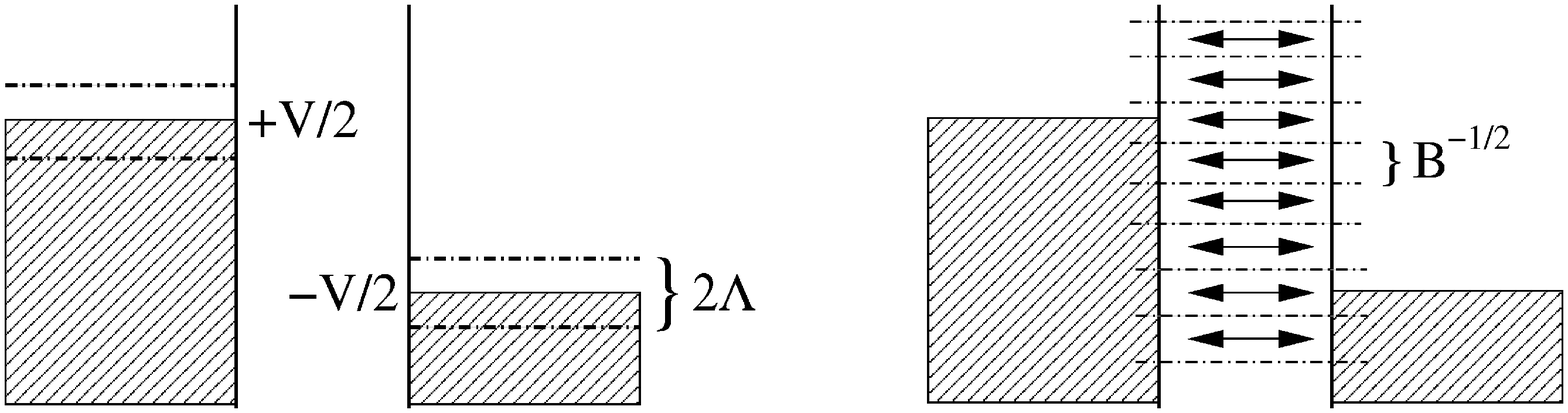}
\caption{Left: Conventional scaling picture where states are integrated
out (here depicted for cutoff~$\Lambda$ smaller than the voltage bias).
Right: Flow equation approach. Here all scattering processes with energy
transfer $|\Delta E|\lesssim B^{-1/2}$ are retained in $H(B)$.}
\label{fig_RG}    
\end{figure}

The flow equation method \cite{Wegner94}
makes a Hamiltonian increasingly diagonal
by applying a sequence of infinitesimal unitary transformations with an 
antihermitean generator~$\eta$: $\frac{dH}{dB}=[\eta(B),H(B)]$. Here $B$ labels
a one--parameter family of unitarily equivalent Hamiltonians and has the dimension
(Energy)${}^{-2}$: $H(B=0)$ is the initial Hamiltonian and $H(B)$ is 
the unitarily equivalent Hamiltonian with matrix elements with energy differences
$|\Delta E|\gtrsim B^{-1/2}$ being eliminated. This 
RG--like separation of energy scales for $B\rightarrow\infty$ can be achieved
by choosing the generator as the commutator of the diagonal and the interaction
part of the Hamiltonian: $\eta(B)\stackrel{\rm def}{=}[H_0(B),H_{\rm int}(B)]$.
In contrast to conventional scaling approaches that successively eliminate
high--energy states in the Hilbert space, the flow equation method keeps all
states but makes the scattering processes increasingly more energy--diagonal.
This method has been successfully applied to numerous equilibrium many--body problems
(e.g., Refs. \cite{Kehrein96,Wegner_Hubbard,Kehrein99}),
where it correctly describes the RG--flow in which the 
UV--cutoff is identified as $\Lambda\propto B^{-1/2}$. 
For an out--of--equilibrium model like the Kondo model with voltage bias,
the difference between state elimination and flow equation diagonalization
turns out to be more fundamental (see Fig.~\ref{fig_RG});
in the flow equation picture scaling below
$B^{-1/2}<V$ is straightforward. 
In particular, the flow equation Hamiltonian $H(B)$ for $B^{-1/2}<V$
still describes the stationary current flowing across the dot since
energy--diagonal scattering processes are not eliminated,  which is
essential for obtaining the current--induced decoherence scale.

We apply the flow equation approach  with the canonical choice of $\eta$
(where $H_0$ is the kinetic energy of the electrons) to the Kondo Hamiltonian
(\ref{Kondo_nonequ}). During the flow higher order interactions are generated
and we parametrize the Hamiltonian as
\bea
H(B)&=&\sum_{t,\alpha} \epsilon^\pdag_t\,c^\dag_{t\alpha} c^\pdag_{t\alpha}
+\sum_{t',t} J^\pdag_{t't}(B)\, \vec S \cdot \vec s^\pdag_{t't} 
\label{HB} \\
&&+i\sum_{t',t,u',u} K^\pdag_{t't,u'u}(B)\,
:\vec S\cdot (\vec s^\pdag_{t't} \times \vec s^\pdag_{u'u}): \nn \ ,
\eea  
where $:\ldots :$ denotes normal--ordering with respect to the
unperturbed Fermi sea \cite{Wegner94}. Here $t',t$ are general indices,
initially $K^\pdag_{t't,u'u}(B=0)=0$ and we neglect newly generated
normal--ordered terms in  $O(J^3)$
and higher. Lengthy but straightforward calculations
lead to the following set of flow equations \cite{Details}:
\bea
\frac{dJ_{t't}}{dB}&=& -(\epsilon_{t'}-\epsilon_t)^2\, J_{t't} 
\label{eq_J} \\
&&+\frac{1}{2}\sum_u J_{t'u} J_{ut} (\epsilon_{t'}+\epsilon_t-2\epsilon_u)
 (n^+(u)-n^-(u)) \nn \\
&&+\sum_{u',u} J_{u'u}\, (K_{u'u,t't}-K_{t't,u'u})\,
n^+(u') n^-(u) \nn \\
&&\qquad\times (2\epsilon_u-2\epsilon_{u'}+\epsilon_t-\epsilon_{t'})
+O(J^4) \nn \\
\lefteqn{\frac{dK_{v'v,w'w}}{dB}=
-(\epsilon_{v'}+\epsilon_{w'}-\epsilon_v-\epsilon_w)^2\, K_{v'v,w'w}} \nn \\
&&\qquad\qquad -J_{v'v} J_{w'w} (\epsilon_{w'}-\epsilon_w) +O(J^3)
\label{eq_K}
\eea
Here the Fermi sea expectation values 
$n^+(u)\stackrel{\rm def}{=}\langle c^\dagger_u c^\pdag_u\rangle$ and 
$n^-(u)\stackrel{\rm def}{=}\langle c^\pdag_u c^\dagger_u\rangle$ arise due 
to the normal--ordering prescription. A nontrivial test for this calculation
is provided by the equilibrium case where $t$ just describes the momentum label.
Using the approximate (however, asymptotically correct in the IR--limit) 
parametrization $\rho J_{p'p}(B)=g(B)\exp(-B(\epsilon_{p'}-\epsilon_p)^2)$ with the
dimensionless IR--coupling constant 
$g(B)=\rho J_{\epsilon_F \epsilon_F}(B)$ in Eqs.~(\ref{eq_J}), (\ref{eq_K}),
one derives the conventional third order
scaling equation for the equilibrium Kondo model, 
$
dg/d\ln\Lambda=-g^2 +g^3/2
$
(with the identification $\ln\Lambda=-(1/2)\ln B$).

Eqs.~(\ref{eq_J}), (\ref{eq_K}) contain complete information 
about the Hamiltonian flow with voltage bias to $O(J^4), O(J^3)$, resp., and can
be analyzed without further approximations \cite{Details}. However, in order
to gain analytical insight one can employ the following approximate
parametrization that focuses on the IR--limit:
$\rho J_{(ap')(ap)}(B)=g_a(B)\exp(-B(\epsilon_{p'}-\epsilon_p)^2)$ with 
$g_a(B)=\rho J_{(a\mu_a)(a\mu_a)}(B)$ for $a=l,r$,
and $\rho J_{(lp')(rp)}(B)=\rho J_{(rp)(lp')}(B)=g_t(B)\exp(-B(\epsilon_{p'}-\epsilon_p)^2)$ with 
$g_t(B)=1/(\mu_l-\mu_r) \int_{\mu_r}^{\mu_l} d\epsilon\,\rho J_{(l\epsilon)(r\epsilon)}(B)$.
Here $g_l$ and $g_r$ are the coupling constants for left--left
and right--right scattering processes located at the respective
Fermi surfaces. For $g_t$ we choose an average over the transport couplings
since this average is directly related to current and
conductance \cite{Approx}. Inserting these parametrizations into
Eqs.~(\ref{eq_J}), (\ref{eq_K}), one arrives at the following set of 
equations that have to be integrated starting from $B=D^{-2}$
($D$ is the initial UV--band cutoff, $a=l,r$):
\bea
\frac{dg_a}{dB}&=&\frac{1}{2B} (g^2_a+g^2_t e^{-2BV^2}) 
-\frac{1}{4B^2}(k_{al}g_l+k_{ar}g_r)
\label{feq_ga} \\
&&-\frac{1}{2B^2}\,
k_{at}g_t \Big(e^{-2BV^2}+V \sqrt{\pi\,B/2}\, 
{\rm erf}(\sqrt{2B}V)\Big) \nn \\
\frac{dg_t}{dB}&=&\frac{1}{2B}\, g_t(g_l+g_r)\,
\frac{\sqrt{\pi}}{2}\,\frac{1}{\sqrt{2B}V} {\rm erf}(\sqrt{2B}V)
\label{feq_gt} \\
&&-\frac{1}{4B^2}(k_{tl}g_l+k_{tr}g_r) \nn \\
&&-\frac{1}{2B^2}\,
k_{tt}g_t \Big(e^{-2BV^2}+V \sqrt{\pi\,B/2}\, 
{\rm erf}(\sqrt{2B}V)\Big) \nn 
\eea
where $dk_{bc}/dB=g_{b} g_c$ with the initial condition
$k_{bc}(B=D^{-2})=0$ for all $b,c=l,r,t$. The scaling picture deduced
from Eqs.~(\ref{feq_ga}), (\ref{feq_gt}) is the main result of this Letter.

\begin{figure}
\includegraphics[clip=true,width=8.0cm]{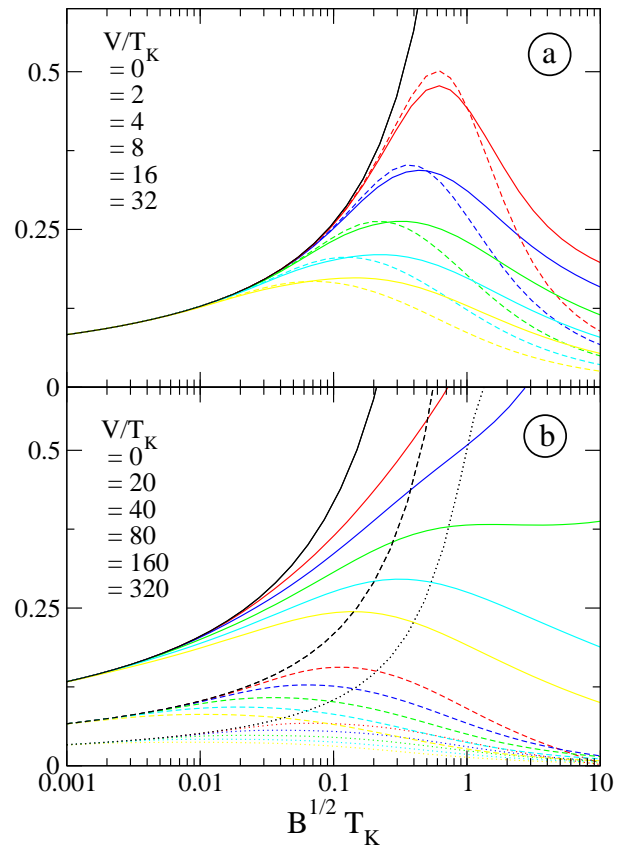}
\caption{Universal curves for the flow of the 
coupling constants $g_l$ (full lines), $g_t$ (dashed
lines) and $g_r$ (dotted lines) for symmetrically coupled quantum dots 
(Plot a with $\Gamma_l/\Gamma_r=1$) and for an asymmetrically coupled
quantum dot (Plot b with $\Gamma_l/\Gamma_r=4$). Results are shown for
various ratios $V/T_{\rm K}$ labelling the resp.\ curves from top
to bottom. $g_l$ and $g_r$ coincide in the symmetric case,
therefore only $g_l$ is shown in a.}
\label{Fig2}    
\end{figure}

Eqs.~(\ref{feq_ga}), (\ref{feq_gt}) take different forms
for $B^{-1/2}\gtrsim V$ and $B^{-1/2}\lesssim V$. We first analyze
the scaling behavior down to the scale set by the voltage bias;
only terms up to second order need to be taken into account here (higher
order terms are unimportant for $V\gg T_{\rm K}$). 
One arrives at a set of equations already analyzed in
Ref.~\cite{Glazman99} ($a=l,r$):
\beq
\frac{dg_a}{dB}=\frac{1}{2B} (g^2_a+g^2_t) \quad , \quad
\frac{dg_t}{dB}=\frac{1}{2B}\, g_t(g_l+g_r) \ .
\eeq 
In the following discussion 
we only look at Kondo dots described by an Anderson impurity model.
The scaling invariant Kondo temperature is then set by the equilibrium 
case, $T_{\rm K}=D\sqrt{g_l+g_r}\exp(-1/g_l+g_r)$, and
one easily shows $g_t(B=V^{-2})=(\sqrt{\Gamma_l \Gamma_r}/\Gamma_l+\Gamma_r)
\,/\ln(V/T_{\rm K})$. From (\ref{feq_gt}) one deduces that the growth of $g_t$
effectively stops below the voltage bias scale since there is no sharp 
Fermi surface for transport processes. However,
looking at the second order terms in (\ref{feq_ga}) the strong--coupling divergences
for $g_l$ and $g_r$ are not cut off. This has led to the prediction
of two--channel Kondo physics in Kondo dots with voltage bias \cite{Coleman01}. 
But even in the weak--coupling regime third order terms in the 
coupling constant 
eventually become more important than second order terms for nonvanishing
voltage bias in (\ref{feq_ga}), (\ref{feq_gt}). For $B^{-1/2}\lesssim V$ 
the dominant terms in the flow equations are approximately
(but sufficiently accurate for a qualitative picture)
\bea
\frac{dg_t}{dB}&=&-\frac{V}{\sqrt{B}}\,\sqrt{\frac{\pi}{8}}\,g_t^3
\label{dec_gt} \\
\frac{dg_a}{dB}&=&\frac{1}{2B} 
\left(g_a^2+2g_a\,\frac{d\ln g_t}{d\ln B}\right) 
\quad {\rm for~}a=l,r \ .
\label{dec_ga}
\eea
The flow changes qualitatively below the {\em decoherence
scale} $B_{\rm dec}^{-1/2}=V\,g_t^2(B=V^{-2})$:
for $B\gtrsim B_{\rm dec}$
algebraic decay $g_t(B)\propto B^{-1/4}$ sets in. In Eq.~(\ref{dec_ga}) one can
then study the competition between coherent strong--coupling behavior
from the second order term, and decoherence effects that arise in linear
order in $g_a$ for $B\gtrsim B_{\rm dec}$. The growth of the coupling constants 
$g_l$, $g_r$ stops
at the decoherence scale unless the coupling constants have already
become too large. This qualitative analysis is confirmed by the numerical
solution of the full set of equations (\ref{feq_ga}), (\ref{feq_gt}) depicted
in Fig.~\ref{Fig2}. Also notice that current--induced decoherence enters
differently from temperature into the flow equation
(\ref{dec_ga}). Temperature acts as an infrared cutoff in the
Kondo strong--coupling terms $g_a^2/2B$, whereas current--induced decoherence
and the coherent strong--coupling processes are {\em in competition}. 
Therefore the Hamiltonian flow derived in the Letter
confirms the existence of the decoherence rate 
$\Gamma_{\rm rel}\propto B_{\rm dec}^{-1/2}$ 
due to current--induced spin relaxation \cite{Glazman99,Rosch}.
It is essential to work in a framework where energy--diagonal
processes are retained because this means 
that a ``window'' of order voltage bias is open for transport 
(compare Fig.~\ref{fig_RG}). It is exactly these transport processes
that are responsible for the emergence of the decoherence scale since they
lead to the $V B^{-1/2}$--terms in (\ref{dec_gt}) and (\ref{dec_ga}).
A conventional scaling approach that removes states around the two
Fermi surfaces and therefore purports to treat energy--diagonal
transport processes (see Fig.~\ref{fig_RG}) cannot describe
this and leads to $1/B$--terms instead 
\cite{footnote_systematic}.

\begin{figure}
\includegraphics[clip=true,width=8.0cm]{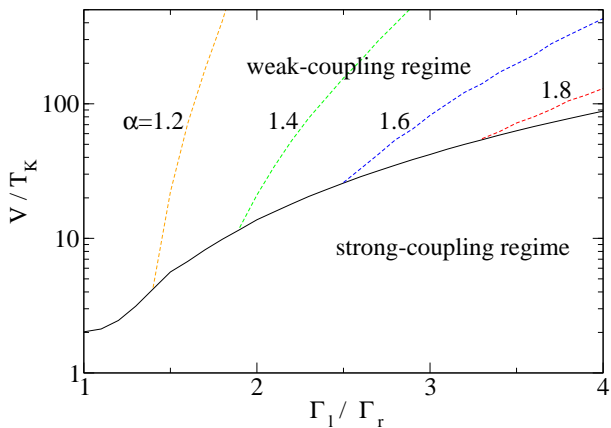}
\caption{Phase diagram of the non--equilibrium Kondo model as a function
of asymmetry $\Gamma_l/\Gamma_r$ and voltage bias. The full line separates
a weak--coupling regime (defined by all couplings remaining smaller than~0.5
during the entire flow) from a strong--coupling regime. The contour lines
(dashed) show the ratio $\alpha=g_l(B_l)\Gamma_r/g_r(B_r)\Gamma_l$, see text
($B_a$ is defined as the $B$--value where $g_a(B_a)$ takes its maximum value).}
\label{Fig3}    
\end{figure}

Since we now know explicitly how decoherence enters into flow
equations (\ref{feq_ga}), (\ref{feq_gt}), 
we can determine a quantitative phase diagram.
The scaling curves for symmetrically coupled Kondo dots in Fig.~\ref{Fig2}a
confirm the absence of two--channel Kondo physics for large voltage bias in 
the sense that all couplings remain small.  We define the 
strong--coupling regime by requiring that at least one coupling grows
larger than~0.5; the resulting curve is depicted in Fig.~\ref{Fig3} \cite{footnote_sc}.
This line should not be interpreted as a phase transition; the
crossover is expected to be smooth --- similar to
the effect of temperature in the equilibrium Kondo model. For asymetrically
coupled Kondo dots this strong--coupling regime extends to remarkably large
values of the voltage bias since the decoherence scale~$B_{\rm dec}^{-1/2}$ is 
proportional to the current~$I$ (see below); for a given
ratio of $V/T_{\rm K}$ the maximum value of $I/T_{\rm K}$ is achieved for symmetrically 
coupled dots (see below). Therefore current--induced decoherence is less effective in
asymmetrically coupled Kondo dots when competing with the coherent 
inter--lead strong--coupling processes.

Experimentally, this phase diagram can be explored by measuring the 
Kondo dot density of states $\rho_d(\omega)$, e.g.\ via a 3--lead setup 
\cite{Lebanon01}. The strong--coupling
regime in asymmetric Kondo dots implies a density of states at the 
Fermi level of the more
strongly coupled lead that only drops significantly ($\lesssim 25\%$)
below the Friedel value once $V/T_{\rm K}$ is well in the
weak--coupling regime in Fig.~\ref{Fig3}. 
For already small
asymmetries the crossover to the strong--coupling regime is driven
by single--channel Kondo physics in the sense that the couplings
at one Fermi surface start to dominate strongly. This effect can
be traced into the weak--coupling
regime by the observation that
the ratio of the local density of states at the two Fermi
levels is then given by $\rho_d(\mu_l)/\rho_d(\mu_r)=\alpha^2$
with $\alpha=g_l(B_l)\Gamma_r/g_r(B_r)\Gamma_l$ from Fig.~\ref{Fig3} 
\cite{footnote_T}. This
observation $\rho_d(\mu_l)/\rho_d(\mu_r)\neq 1$
in asymmetric Kondo dots has also been made in other
approaches though without obtaining a quantitative phase diagram
(compare Refs.~\cite{Rosch,Lebanon01,Krawiec}).

A final remark regarding the calculation of observables:
these need to be unitarily transformed as well
\cite{Kehrein96} (they typically change their form completely
once $B\gtrsim B_{\rm dec}$).
For example  the current
operator remains form--invariant with
the flowing coupling $g_t(B)$ up to scale $B_{\rm dec}$:
$
I(B)=ig_t(B) \sum_{p',p} \vec S\,\cdot\,
(\vec s_{(lp')(rp)} - \vec s_{(rp')(lp)}) 
$.
One can then work with the renormalized effective
Hamiltonian on the scale $B_{\rm dec}$ and a conventional
Keldysh calculation yields $I=(3\pi/4)V\,g_t^2(B_{\rm dec})$.
This leads to the well--known perturbative result for the conductance 
\cite{Glazman99}
$
G(V)=G_{\rm u}(3\pi^2/16)/\ln^2(V/T_{\rm K})
$
where $G_{\rm u}=(e^2/\pi\hbar)\,4\Gamma_l\Gamma_r/(\Gamma_l+\Gamma_r)^2$
is the conductance in the unitarity limit. Notice that transport quantities
are not low--energy properties like the Kondo dot density of states 
at the Fermi levels and are therefore unaffected by the strong--coupling
physics in Fig.~\ref{Fig3} as long as $V\gg T_{\rm K}$.

Summing up, we have developed a Hamiltonian scaling picture for
Kondo dots with voltage bias that allows us  to express
physical quantities in terms of renormalized parameters. We confirm the
absence of two--channel Kondo physics for symmetrically coupled quantum
dots, and show the existence of a large regime dominated by 
single--channel strong--coupling
physics for asymmetric dots. The flow equation approach
presented here should be useful for other non--equilibrium models as well.

I thank J.~von Delft, D.~Vollhardt and F.~Wegner for valuable discussions,
and acknowledge valuable discussions about this and related subjects with 
N.~Andrei and B.~Doyon.
This work was supported in part by SFB~484 of the Deutsche 
Forschungsgemeinschaft (DFG) and by a Heisenberg fellowship 
of the DFG.


\begin{thebibliography}{11}
\bibitem{Details}
Details of this calculation will be published separately.
\bibitem{Approx}
This approximation is not always accurate, e.g.\ for
describing the flow of $J_{(l\mu_r)(l\mu_r)}$ or $J_{(l\mu_r)(r\mu_r)}$
for $B^{-1/2}\lesssim V$.
However, parametrizations with more coupling constants 
do not modify our conclusions and we work with the minimum
number necessary.
\bibitem{footnote_systematic}
One can also verify that higher
order terms beyond the order taken into account in (\ref{feq_ga}),
(\ref{feq_gt}) are smaller than the terms calculated here
as long as all the coupling constants remain sufficiently small
during the entire flow.
\bibitem{footnote_sc} This crossover line only depends weakly on the specific
value for the running coupling constant that determines the strong--coupling
regime (here 0.5), it is e.g.\ hardly changed for a value of~1.0. Also
higher orders in the calculation can shift this line, though
the third order results shown here are expected to be a reliable approximation.
\bibitem{footnote_T}
Close to the crossover line the number $\alpha^2$ is a lower bound for the asymmetry:
the real ratio will be larger since one also needs to 
consider the $K$--couplings
in (\ref{HB}) \cite{Details}.
\end{thebibliography}
\end{document}